\documentstyle[12pt,psfig]{article}
\def\bold#1{\setbox0=\hbox{$#1$}%
      \kern-.025em\copy0\kern-\wd0
      \kern.05em\copy0\kern-\wd0
      \kern-.025em\raise.0433em\box0 }

\def\eea{\end{eqnarray}}
\def\bea{\begin{eqnarray}}
\def\eeas{\end{eqnarray*}}
\def\beas{\begin{eqnarray*}}
\def\ee{\end{equation}}
\def\be{\begin{equation}}
\def\bdm{\begin{displaymath}}
\def\edm{\end{displaymath}}

\begin{document}

\begin{titlepage}
\begin{center}
\vspace*{2.0cm}
{\Large\bf Quantum spectra of triangular billiards on the sphere }
\vskip 1.5cm

{M.E.Spina and M.Saraceno}
\vskip .2cm
{\it Department of Physics, \\ Comisi\'on Nacional de Energ\'{\i}a At\'omica,
          Av. Libertador 8250, \\
(1429) Buenos Aires, Argentina. }
\vskip 2.cm
November 2000 \\
\vskip 2.cm
{\bf ABSTRACT}\\
\begin{quotation}
\noindent We study the quantal energy spectrum of triangular billiards on a spherical surface.
Group theory yields analytical results for tiling billiards while the generic case is treated numerically. We find that the statistical properties of the spectra do not follow the standard random matrix results and their peculiar behaviour can be related to the corresponding classical phase space structure.
\end{quotation}
\end{center}
\vspace{1cm}

\end{titlepage}

\section{INTRODUCTION}

\noindent The classical dynamics in polygonal billiards  (in particular equilateral triangles) on a spherical
surface  was studied in \cite{spina} to investigate the effect of the positive curvature on the classical motion.
The structure of phase space for these curved triangular billiards turned out to be regular but very complex. As a
consequence of the focusing mechanism on the sphere the phase space  is entirely covered by chains of  stable
islands. Inside each of these islands the motion is elliptic and can be labelled by an infinite repeating code
according to the sequence of reflections.  This  situation  that we could call   'piecewise integrable'  is very
different from the one corresponding to plane polygonal billiards which are either integrable (or pseudointegrable)
or ergodic, according to the rationality of their inner angles (see, for example, \cite{gut}). \noindent The
natural question we would like to answer in the present work is how this peculiar dynamics is reflected in the
statistical properties of the quantum mechanical spectrum. 

\noindent There is a lot of evidence, mainly numerical, that
integrable dynamics leads to Poisson statistics while classically chaotic systems satisfy  the statistics given by
one of the Random Matrix Ensembles (see, e.g., \cite{bohi}).  Mixed systems show a combination of these extremes
that depend on the relative size of chaotic and regular regions \cite{berob}. On the other hand, a different kind
of systems neither regular nor chaotic has been shown to conform to intermediate statistics: a simple example is
the plane pseudointegrable billiard which has been studied together with other models with similar behaviour by
Bogomolny et al \cite{bogo}. 

\noindent The peculiar phase space structure of equilateral triangles on the sphere led us to believe that
their spectral properties would not conform to any universal statistics. In this article we calculate the quantal energy spectrum and the eigenfunctions of these systems.
 For generic triangles the
calculation  will be performed numerically while the spectra of tiling triangles will be derived analytically just
by using symmetry arguments, as was the case in the classical counterpart.  It was shown in \cite{spina} that
tiling triangles are very particular systems that not only are integrable but for which only periodic orbits are
present. We will see how this peculiarity affects the quantal spectrum. \noindent In section 2 we present the model
and the numerical procedure to calculate the spectra.  Section 3 is devoted to the spectral properties of tiling
triangles. In section 4 we derive the level spacing distributions of several generic triangles and discuss them in
connection with the corresponding classical phase space. Finally conclusions are presented in section 5.

\section{THE MODEL}

\noindent Given $ T $ an equilateral triangle with inner angle $ \omega $ centered on the north pole of a sphere of
radius $ R =1 $, we are concerned with the eigenvalue problem

\be
\triangle  \psi_E + E \psi_E =0   \qquad in  \quad T  \qquad and  \qquad   \psi_E = 0   \qquad    on  \quad  T
\label{hel}
\ee

\noindent where $ \triangle $ is the three-dimensional Laplacian in spherical coordinates
$ \theta $ and $ \phi $ given by:

\be
\triangle = {1 \over \sin \theta}  {\partial  \over { \partial \theta}}  \quad ( \sin  \theta  {\partial
\over { \partial  \theta} })  +{1  \over  \sin^2 \theta}  \quad  {\partial ^2  \over  {\partial \phi^2}} .
\label{lap}
\ee

\noindent  and the eigenvalues $ E  = { {2 m}  \over \hbar }  \cal E $,  $ \cal E $ being the energies,
 are dimensionless.

\noindent The wave function $ \psi_E $ can be expanded as

\be
 \psi_E =  \displaystyle \sum_m   c_m  F^m (E, \cos  \theta)   e^{i m \phi}
\label{exp}
\ee

\noindent  where $ m $ is an integer and  $ F ^m $ is a solution of the  Legendre's equation:

\be
{ \partial \over { \partial x}} \quad [ (1 - x^2) F (x) ] \quad + \quad [{ -m^2 \over (1 - x^2)}
+ E ]  \quad  F(x) \quad = \quad 0
\label{lege}
\ee

\noindent with  $ x = \cos \theta $.
The functions $ F^m $ which have to be regular inside the triangle are then proportional to the Legendre's
functions  of the first kind $ P_{ \nu} ^m (x) $, with real $ \nu $ such that $ E = \nu  ( \nu + 1) $ and real
argument  $ -1 < x  < 1 $.  These $ P_{ \nu} ^m $ ( $ m  \neq  0 $) go to $ 0 $ at  $ x = 1 $, while they diverge
for $ x = -1 $.

\noindent It will be convenient to define :

\be
 F^m (E, x ) = \sqrt { \Gamma ( \nu - m + 1) \over \Gamma ( \nu + m + 1) }  \quad
 P_{ \nu} ^m (x)
\label{norm}
\ee

\noindent
These functions are real for $ m \leq \nu + 1 $ and satisfy the following recursion relation:

\begin{eqnarray}
F^{m+2} & = & - \sqrt{ 1 \over { (\nu+m+2) (\nu-m-1)} } 2 (m+1) { x \over { \sqrt{1 - x^2}} }\quad
F^{m+1}  \nonumber \\
  &  & - \sqrt { {(\nu-m) (\nu+m+1)} \over {(\nu+m+2)(\nu-m-1)}}\quad F^m
\label{recur}
\end{eqnarray}

\noindent which is used in the numerical procedure to compute the $ F^m $.

\noindent
In order to obtain a spectrum free from the trivial degeneracies due to symmetry we have to desymmetrize the
billiard. The spherical equilateral triangle is invariant under the point group $ C_{3v} $ and can be cut along the
$ 3 $ reflection planes into $ 6 $ triangular subdomains. Solving the eigenvalue problem in each of these
irreducible domains with given boundary conditions corresponds to finding a fraction of the spectrum of a given
symmetry class. In the following we will treat the Dirichlet problem in the fundamental triangle. This corresponds
to solutions which are odd under reflection across all the symmetry planes. Subspectra corresponding to other
symmetry classes can be obtained by imposing Neumann and mixed boundary conditions. In the case we are considering
the expansion of Eq.(\ref{exp}) reduces
to:

\be
 \psi_E =  \displaystyle   \sum_p   c_{3p}   F^{3 p} (E, \cos  \theta)   \sin  3  p \phi
\label{exp3}
\ee

\noindent that vanishes automatically at $ \phi =0 $ and $ \phi = { \pi  \over 3} $.

\noindent In order to solve the problem we use an improved version of the 'point matching' method
introduced in \cite{schmi}. This  consists in requiring that the Fourier coefficients of the wave
function evaluated on the boundary

\be
I_n = \int ds e^{-2 i \pi n s / \cal L}  \psi_E
\label{int}
\ee

\noindent should vanish.  $\cal L $ is the billiard perimeter and $ ds $ the line element.
The existence of a non trivial solution  leads to the condition:

\be
det [ J_{n,m} ] = 0, \qquad  with \quad n,m = 1 ,  N
\label{det}
\ee

\noindent where
\be
J_{n,m} ( E )= \int ds  e^{-2 i \pi n s / \cal L}   F^m (E, \cos \theta(s) )  e^{i m \phi(s)}
\ee

\noindent Eq. (\ref{det}) determines the energy levels $ E $.

\noindent  The numerical calculation was  performed  for equilateral triangles with inner angle $ {  \pi \over 2}
\leq  \omega  \leq  { 2 \pi \over 3}$, that is fairly large triangles for which no difficulties are expected at the
corners. Before presenting these numerical results we analyze in the next section the particular cases of tiling
triangles that can be treated analytically.

\section{TILING TRIANGLES}

\noindent As already described in \cite{spina} the classical motion in tiling triangles can be studied by following
a unique geodesic on the topological surface obtained by sewing together a finite number of replica of the original
billiard. Since the geodesics are closed curves on this compact surface, all orbits are periodic.  In other words,
every trajectory in a tiling triangle is restricted to a one dimensional subspace. The system is over-integrable,
the compactness of the sphere playing the role of an additional integral. We therefore expect a non generic
behavior of the level distribution. \noindent The eigenvalues and eigenfunctions of tiling triangles can be
evaluated  by making use  of symmetry arguments exclusively. Let us consider the equilateral triangle with  $
\omega = {\pi \over 2}$, that is, a triangle whose vertices coincide with those of a face of an octahedron.
Following the desymmetrisation scheme of the previous section we cut the triangle into $6$ triangles with Dirichlet
boundary conditions and calculate the 'desymmetrised spectrum'. \noindent Since the domain tesselates the sphere
the solutions of  Legendre's equation (\ref{lege}) have to be continuous and one-valued at all points  $ -1 \leq
\cos \theta  \leq  1 $. This restricts  $ \nu $ to integer values $ l $ and $ |m| \leq l $. The seeked
eigenfunctions will then be linear combinations of  spherical harmonics $ Y_l^m$ that can be determined by
requiring that they should vanish on the boundaries. These boundary conditions result in a reduction of  the
spherical symmetry of the problem: the full spherical symmetry group including all proper and improper rotations in
three dimensions will then be reduced to a point group. For this particular triangle it will be the symmetry group
of the octahedron . That means that in order to determine the eigenvalues spectrum, that is, the allowed values of
$ l $ and their degeneracy $ \lambda_l $ we should determine how each of the representations of the full spherical
group $ D_l^{\pm} $ may be decomposed into  irreducible representations of $ O_h $.

\noindent
The number of times the $ \alpha ^{th} $ irreducible representation of the subgroup $ O_h $ is contained
in the representation $ D_l^{\pm} $ of the spherical group is given by \cite{table}:
\be
c_{\alpha} = {1 \over g_{O_h}}   \displaystyle \sum_R \chi*_{ \alpha} (R)   \chi_l (R)
\label {decom}
\ee

\noindent where the sum extends over the elements of the group $ O_h $ the order of
which is $ g_{O_h} = 24 $. The characters $ \chi_{ \alpha} (R) $ can be extracted
from the corresponding character table while the characters for the proper rotations
in the full rotation group are given by \cite{table} :

\be
\chi [D_l^{\pm} ( R_{\theta})] = {{\sin ( l+1/2) \theta} \over {\sin {\theta \over 2}}}
\label{fullr}
\ee

\noindent Here  $ R_{\theta}$ denotes a rotation through  $ \theta $ about some axis. The character of an improper
rotation (that is, proper rotation times an inversion) is the same as the character for the proper rotation  for
the $ + $ representations (that is, for even values of $ l $) and its negative for the $ - $ (odd values of $ l $)
representation.

\noindent By applying  he $ 24 $ symmetry operations of $ O_h $ it can be
easily seen that the eigenfunctions satisfying Dirichlet conditions in the
reduced triangle transform as the one-dimensional irreducible
representation $ \Gamma_1^- $  of $ O_h $ . Therefore the degeneracy
$\lambda_l $ of the eigenstate  of energy $ E_l = l (l+1) $ will be the
coefficient of the representation $ \Gamma_1^- $ in the decomposition of $
D_l^{\pm} $ according to Eq.( \ref{decom}).

\noindent The first levels of the spectrum are
shown in Table 1  with their corresponding
degeneracies. The eigenfunctions have been
obtained  by means of diagonalizing the projector
of the representation $ \Gamma_1^- $  of $ O_h $ :
they  are the eigenfunctions corresponding to
eigenvalues equal to 1. Since the representation
is one-dimensional the projector has been
expressed in terms of the characters as:
\be
P_{ \Gamma_1^-} = {1 \over g_{O_h}}   \displaystyle \sum_R  \chi_{ \Gamma_1^-} (R)  D_l^{\pm} (R)
\label{proj}
\ee

\noindent  Again the sum runs over all the elements of the group, and the  D-matrices corresponding to improper rotations are minus the corresponding to proper rotations.

\noindent It is easy to understand that only odd values of $ l $ appear in the spectrum of Table 1
since the representation $ \Gamma_1^- $ appears only in the decomposition of
$ D $  representations with negative (natural) parity.  Furthermore,  it can be seen that for $ l \geq 13 $ all odd values of $ l  $ are present in the sequence. That means that except for the lowest lying levels the level spacing $ S(l) = E_{l+2} - E_l $  is proportional to $ l $.  In order to keep constant the classical mean level spacing  $ \bar S $ which  according to Weyl's law  is  $ \bar S = {4 \pi \over A} $, $ A $ being the area of the reduced triangle, the level degeneracy $ \lambda_l $ should also increase linearly with $ l $.  The spectrum of the tiling spherical triangle is thus dominated by number-theoretic degeneracies.
These 'accidental' degeneracies have been studied by Itzykson et al \cite{itzy}. They appear in some simple integrable quantum systems like, for instance, the harmonic oscillator with rational frequency ratios ( where $ E_{m,n} = m+n+1 $ ) and  the plane integrable polygonal billiards  ( for the equilateral triangle $ E_{m,n} = m ^2 +n^2  -m n $ ). The harmonic potential  was extensively investigated by Berry and Tabor  \cite{berryta} . It is an example of overintegrable systems, in which the additional integrals are the conmensurability relations. The spacing among adjacent levels is constant while the degeneracy increases with the energy . This makes the mean level spacing go to $ 0 $ in the classical limit and no spacing distribution can be defined.  The case of the plane equilateral triangle presents more  analogy with our case in what concerns the mean level spacing   $ \bar S $  which is  well defined and given by Weyl's formula.  Also there the states are increasingly degenerate and separated by increasingly large gaps.

\noindent
In order to study the statistics of the nearest neighbor spacing distribution $ P (S) $ we now unfold
the spectrum by defining the sequence $ e_l = \bar N ( E_l )  $ where $ \bar N (E) $ is the averaged integrated
level density given by

\be
 \bar N(E) = {A \over 4 \pi} E - {L \over 4 \pi} \sqrt E + const
\label{weyl}
\ee

\noindent $ L $ is the triangle perimeter in the spherical metric. The level spacing distribution  in the scaled
spectrum $ P(s) $, with $ s_l =e_{l+2} - e_l $ is plotted in Fig.1 (b)  for a sequence of $ 7500 $ levels.  The
distribution is bimodal with  a strong tight peak at $ s = 0 $ and  a flat component, giving equal probability to
all allowed values of $s _l  $.  Since both degeneracies and gaps diverge with the energy this distribution is not
defined in the classical limit.  Asymptotically the peak tends to a $ \delta $- function and the flat mode extends
to infinite values of $ s $ with a height going to $ 0 $. For comparison  in Fig.1 (a)  we show the  spacing
distribution for a sequence of $ 20000 $ energy levels corresponding to a desymmetrized plane equilateral triangle
with Dirichlet boundary conditions . Although the distribution looks more poissonian, as in a generic regular
system, the situation is analogous to the one in the curved triangle: as shown in \cite{pinsky} degeneracies and
gaps diverge with energy. But, as pointed out by Berry in \cite{berrytri} for the similar case of the right plane
triangle, this divergence is so slow that  this nongeneric level structure governed by number-theoretic
degeneracies only appears for very high-lying states. \noindent Summarizing,  the spectrum of tiling spherical
triangles is non generic and although a mean level spacing  can be defined, no spacing distribution $ P(s) $ exists
in the classical case. This peculiarity is to be attributed to the number-theoretic structure of the spectrum .
\noindent  The spectra of the spherical tiling triangles with angles  $ \omega = { 2 \pi \over 3}$ and  $ \omega =
{ 2 \pi \over 5}$ can be calculated in the same way, by considering the symmetry group of the tetrahedron and of
the icosahedron respectively. \noindent The sequence of eigenvalues corresponding to the desymmetrized triangle
with   $ \omega = { 2 \pi \over 3}$ and Dirichlet boundary conditions, that transform as the one-dimensional
irreducible representation $ \Gamma_2$  of $ T_d $ is presented in Table 1. In this spectrum odd and even values
of $ l $ are allowed . Except for the lowest lying levels all values of $ l $ are present, the level spacing  $
S(l) = E_{l+1} - E_l $ is again proportional to $ l $ and the spectrum has the general characteristics stated
above.

\section{GENERIC TRIANGLES}

\noindent
The  numerical procedure introduced in section 2 was carried out for curved equilateral triangles with
inner angle $ \omega $ ranging from   $ {  \pi \over 2} $  to $  { 2 \pi \over 3}$. We checked the stability of the
solutions by varying the number of points on the boundary used to evaluate the integral in Eq.(\ref{int}) and the
number of partial waves in the expansion Eq.(\ref{exp}). The completeness of each spectrum was tested by comparing
the calculated cumulative level density $ N(E)$ with the averaged one given by the Weyl formula (\ref{weyl}).  This
test is particularly relevant for triangles close to the tiling ones,  for which most of the levels are
near-degenerate and therefore some of them could be missed in the numerical calculation.

\noindent In Fig. 2  we
show the energy spectra up to $ E = 2400 $ for a family of equilateral triangles  labeled by $ \omega $. These are
plotted as curves  $ E_{\nu} (\omega) $. The spectra at the left and right boundaries correspond to the integrable
triangles with $ \omega =   {\pi \over 2} $ and  $\omega = { 2 \pi \over 3}$ respectively, which have been
evaluated analytically as described in the previous section. The levels in both sequences are labelled by an
integer $ l $ and their multiplicity increase with energy.

\noindent All the remaining triangles with arbitrary
rational or irrational  angle $ \omega $ are non-integrable and therefore present no degeneracies in their spectra.
Accidental degeneracies are not expected neither: by varying  $ \omega $, we are moving in a one-parameter family
and, according to Berry \cite{berrydia}  and references therein, variations in one parameter are insufficient to
produce degeneracies.  However, in the region denoted as $ III $ we observe a great number of quasi-crossings. Their
origin will be explained below.

\noindent The level spacing distributions corresponding  to the unfolded spectra of
six triangles with $ \omega $ ranging from  $ {\pi \over 2} $ to  $ { 2 \pi \over 3}$  are shown in Fig. 3. Between
$ 1300 $ and $ 2500 $ levels were considered for each case, except for case (d) ($ \omega = 0.589 \pi $) where the
huge number of quasidegeneracies prevented us to go beyond a few hundreds levels. We observe that the distributions
depend drastically on the parameter $ \omega $ and do not exhibit any universal behaviour. In order to understand
the general features of the distributions and their angle dependence we will have to keep in mind the results for
tiling triangles derived in the previous section and also refer to the classical phase space plot which is shown
for each case  in Fig. 4. As seen in  \cite{spina} for generic curved triangles the classical phase space is
covered by chains of elliptic islands of regular motion characterized by an infinite repeating code. The
multiplicity of these chains increases and their size decreases with the period of the code and phase space takes a
fractal structure. As shown in Fig. 4 (a) and (f), for triangles close to the tiling ones phase space is almost
dominated by one chain of  $ 3 $ islands (corresponding to the $ +++$ code ) in the case  $ \omega = {\pi \over 2}
$, and by two chains (corresponding to $ +++$  and  $ +-$ ) in the case of  $ \omega = {2 \pi \over 3} $.  The
region outside these big domains is entirely covered by 'dust', that is  by chains of islands corresponding to very
long  codes and therefore with extremely high multiplicity and small area. These islands , although regular and of
the same type as the dominant ones, cannot be resolved quantum mechanically in the region of the spectrum we are
analyzing. As we go to triangles far apart from the tiling ones (see Fig. 4 (b),(c), (d),(e)) other chains of
considerable size appear and we will see that some of them can be resolved in the range of energies considered.
\noindent In order to see how this is reflected in the level spacing distributions we now go back to Fig. 3.

\noindent Histogram $(a)$ corresponds to an angle $ \omega = 0.511  \pi $  ( region $ I $ of Fig. 2) and can be seen as a
perturbation of the tiling    $ \omega =  {\pi \over 2} $.  The $O_h$  symmetry is broken and the  fact that
degeneracies disappear and  gaps between levels do not diverge with energy but tend to some finite value reflects
in a modification of the distribution of  Fig. 1 (a).  The  distribution  Fig. 3 (a)  is still  bimodal but the
delta peak at $ s=0 $ is shifted (i.e., there is some level repulsion)  and the flat component does not extend to
large values of $ s $. Case  (f)  corresponding to  $ \omega = 0.649  \pi $ (also in region $ I $ of Fig. 2 )
presents an analogous situation, with a breaking of the $ T_d $ symmetry. But in this histogram the low lying peak
is broader (in fact, there is no level repulsion) and the high s - component shows more structure suggesting a
superposition of two uncorrelated spectra. In both cases the distributions, and more specifically the presence or
not of level repulsion, can be better understood by analyzing the corresponding eigenfunctions.  At a qualitative
level, we expect that eigenstates spanning an integrable region will show rigid and regularly spaced spectra.
Therefore, in case where a single integrable region dominates the phase space the spectrum will show repulsion,
since all the eigenfunctions are correlated. On the contrary,  if several classes of eigenfunctions coexist, the
distribution can be thought as a superposition of uncorrelated spectra and there will be no level repulsion. In
case (a) the eigenfunctions are of two types: one class lives in the $ +++ $ elliptic islands which mostly fill the
phase space,  the other class extends over the region left outside the islands, which is a very small fraction of
the space.  (see Fig. 4 (a) ). Thus, the eigenfunctions of the first class, which are the majority, will be
correlated leading the  level repulsion observed in Fig. 3 (a).
In case (f) the eigenfuctions are of
three types. Apart from the extended class which lives in  the region left outside the islands, there are two
classes of  localized functions: one living in the $ +++ $ elliptic islands, the other in the  $ + - $ islands
(see Fig. 4 (f) ).  Examples of the Husimi representation of functions belonging to these three classes are shown
in Fig. 5. The two classes of localized eigenfunctions living in domains which are comparable in size will be
weakly correlated : this explains the much weaker level repulsion in distribution of Fig. 3 (f). 

\noindent The
distributions corresponding to $ \omega = 0.525 \pi $ and $ \omega = 0.633 \pi $ are shown in Fig. 3 (b)  and (d)
respectively. Both correspond to  values of the parameter in region $ II $ of Fig. 2 .  We are now fairly far away
from the tiling triangle: the splitting among levels is big, and the two modes have shifted yielding a single
peaked distribution.  As we  see in the  corresponding classical phase space plot Fig. 4 (b)  for  $ \omega = 0.525
\pi $ the islands corresponding to codes different from the dominating   $+++$ are still small and cannot be
resolved in the range of energies we are considering.  Still, there are  two classes of eigenstates  coexisting:
the ones localized in the $+++$ domain and the extended ones which occupy now a considerable fraction of phase
space. Therefore two classes of weakly correlated eigenfunctions contribute to the spectrum.  For $ \omega = 0.633
\pi $ three classes of functions are present, since the $ +- $ islands  are large, as shown in Fig. 4 (d), and can
be resolved.

\noindent Summarizing, we can say that the spectra of generic curved triangles are non universal and
follow a parameter- dependent  intermediate statistics.  The analysis of the corresponding classical phase space
plots  suggest that the characteristics of the distributions  depend on the number of domains that can be quantum
mechanically resolved in a given energy range.  The conjecture is the following. If only one big domain is resolved
the eigenfunctions in this domain are of the same type and, therefore, correlated. If more than one domain can be
resolved, the coupling among states living in different domains will be weak, and the spectrum will be a
superposition of uncorrelated spectra, showing  no repulsion. 

\noindent Another interesting observation is the
existence of a high peak at a small value of $ s $ in the level spacing distributions corresponding to triangles in
region $ III $ of Fig. 2. Two examples are shown in In Fig.3  (c) and (d) for angles  $ \omega = 0.571 \pi $ and
$\omega = 0.589 \pi $ respectively. This peak is due to the presence of numerous quasi-crossings in this region, the
lowest of them are seen already in  Fig.2. To understand the origin of these quasi-degeneracies we have to refer
once again to the classical  phase space plots corresponding to these triangles.  As pointed out in \cite{spina}
generic non tiling triangles might have domains in which all trajectories are periodic (coexisting with the
families of elliptic islands). For example, in all rational triangles with   $ \omega > {\pi \over 2} $ the islands
corresponding to the $ +- $ code are constituted of periodic orbits.  On the other hand,  in triangles with inner
angle satisfying

\be
\cos { \omega \over 2} = \sqrt{ {3 \over 4} - \cos ^2 { k \pi \over n}}
\label{titi}
\ee

\noindent all orbits of type $ +++ $ are periodic with period $ n $.

\noindent In region $ III $ we find several  triangles with rational inner angle or inner angle satisfying Eq.
(\ref{titi}), corresponding to periodic orbits of relatively short period. For example, in case (d)  $ \omega $
satisfies Eq. (\ref{titi}) with period $ n = 7 $.  Although there is no global symmetry group, as in the case of
tiling triangles, the presence of these periodic domains which cover a big fraction of phase space (in particular
the corresponding to the $ +++ $ code) lead to quasi degeneracies among the quantum states localized in these
domains. This is the origin of the huge peak at $ s = 0 $ in Fig. 3 (d). Distribution (d) corresponds to an angle
close to $ {4 \pi \over 7} $ : now the periodic orbits are of type $ +- $ with period $ n = 7 $. Since the area
occupied by the $ +- $ islands is smaller, the effect is less pronounced: the peak is lower and located at $ s $
small but different from $ 0 $.

\section{CONCLUSIONS}

\noindent  In this work we calculated the quantal spectra of generic and tiling equilateral triangles on a
spherical surface.  The spectra of the tiling triangles could be obtained analytically by using symmetry arguments
while for the generic ones we performed a numerical calculation based on an improved version of the 'point
matching' method which allowed us to go up to around $ 2000 $ levels.

\noindent We found that the spectra of tiling triangles are dominated by arithmetical degeneracies. This is a
property shared by other simple quantum systems, in particular by plane square and triangular billiards. In the
case of curved triangles the distribution of the degeneracies is easily determined: since the level gap increases
linearly with $ l $, that is,  is proportional to $ \sqrt E $ and the mean level spacing in the classical limit is
constant, the degeneracies also increase as  $ \sqrt E $. But the question asked in \cite{itzy} about relating
these degeneracies with some 'hidden' symmetry remains open also in this case.

\noindent As expected, the spectra of generic triangles of variable inner angle $ \omega $ show no universal
behavior in this energy range.  We first remarked that for triangles close to the tiling ones the level spacing
distribution is bimodal : this can be seen as a remnant of the level clustering in the tiling systems. More in
general, for an interpretation of the spectral properties we had to refer to the corresponding classical phase
space plots. These might be governed by one or more than one island of considerable size, where considerable means
that it can be quantum mechanically resolved in the energy range under consideration.  The conclusions seem to be
the following. If there is only one resolved domain all the eigenfunctions living in this domain will have support
on tori and the corresponding energy levels will be regularly spaced, giving rise to a rigid spectrum.  This is the
case in triangles with inner angle $ \omega $ close to $ \pi \over 2 $ for which the $ +++ $ islands covers a great
area of the phase phase ((a) and (b) in Fig.4).  If several domains of comparable size coexist different
eigenfunctions which may correspond to levels close in energy are localized far apart from each other and are
weakly correlated.  No level repulsion will be present in the spectrum.

\noindent In order to check these conjectures and to get a quantitative understanding of the spectra we have to
extend the calculation to higher energies. The higher we go, more and more domains in phase space will be resolved
and an increasing number of weakly correlated families of eigenfunctions will be present, localized in these
domains. It will then be possible to subdivide the spectrum in subspectra corresponding to each type of
eigenfunctions (according to their localization in phase space) and study the characteristics of each subspectrum.
This subdivision is not possible in the present work since we cannot reach values of $ k $ for which the small
islands will be explored with significant statistics. To remedy this situation  and check whether in the high $ k $
regime the superposition of many uncorrelated spectra  recovers asymptotically the Poisson behaviour,  we are at
present adapting the scaling method  \cite{scaling} to the geometry of the sphere.

\vskip 2.cm
\subparagraph{Acknowledgements}
\noindent We are grateful to N.N. Scoccola for valuable 
discussions. This work was partially supported by PICT97 03-00050-01015 
and CONICET PIP 0420/98.

\pagebreak

\pagebreak

\vskip .1cm
\centerline{ { \psfig{figure=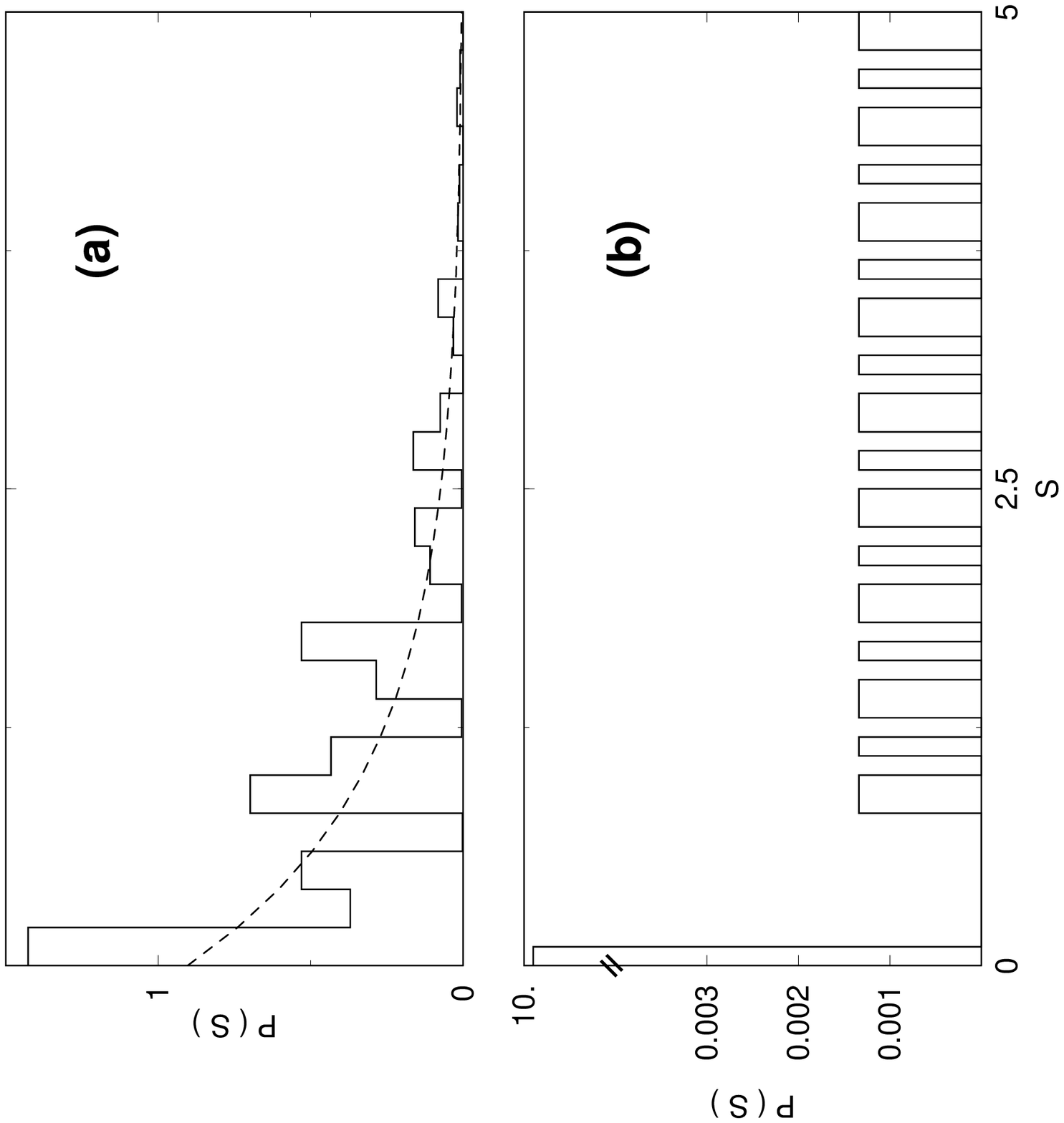,width=20.cm,height=15.cm,angle=0} } }
\vskip .5cm
\centerline{ \parbox{15cm}
{{\normalsize {Fig.1 -- (a) shows the level spacing distribution corresponding
to the first $ 20000$ levels of a desymmetrized plane equilateral triangle.
In (b) the level spacing distribution for the first $ 7500 $ levels of a desymmetrized
curved  tiling triangle with $ \omega = { \pi \over 2} $.} }} }

\pagebreak

\vskip .1cm
\centerline{ { \psfig{figure=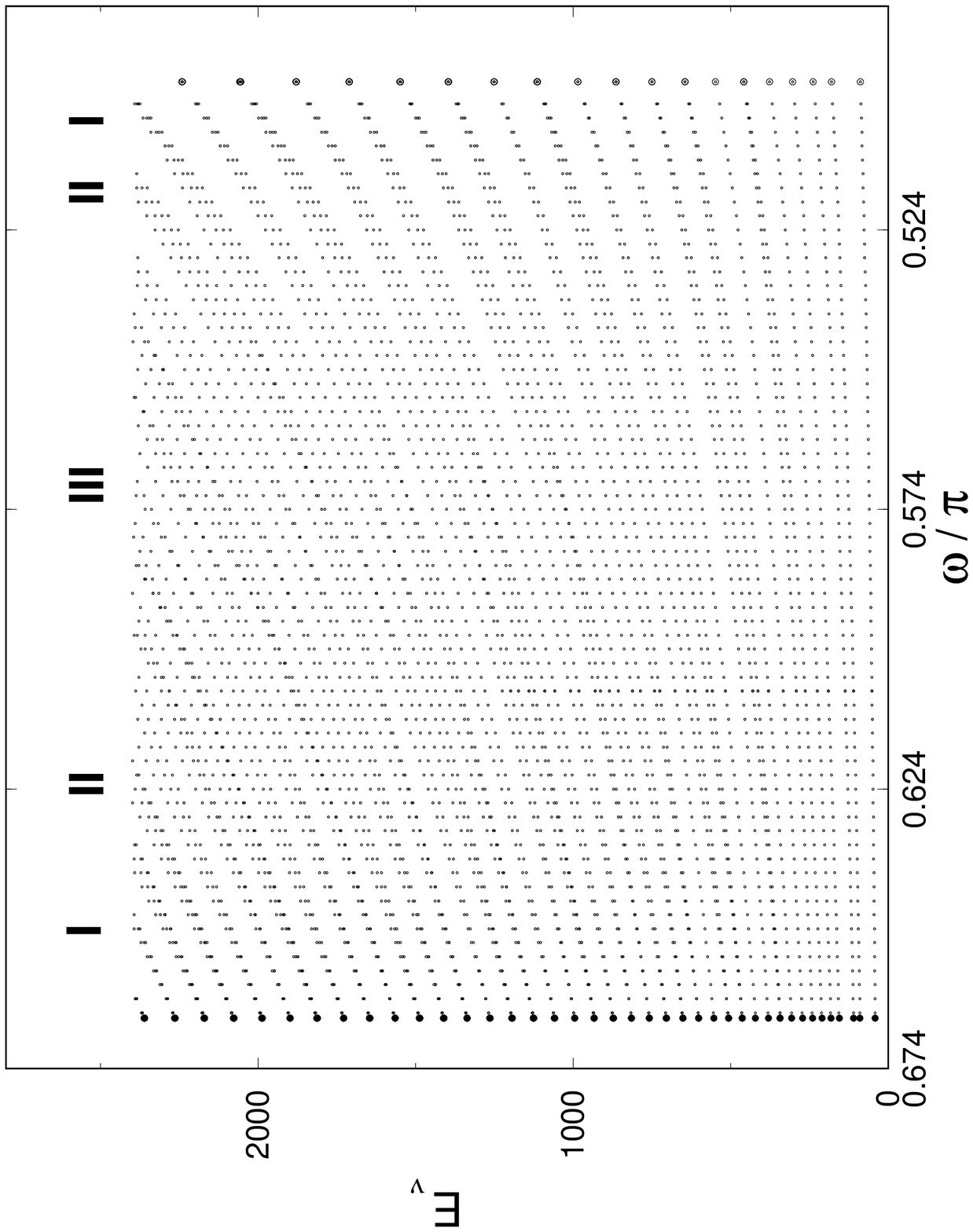,width=20.cm,height=20.cm,angle=0} } }
\vskip .5cm
\centerline{ \parbox{15cm}
{{\normalsize {Fig.2  Energy spectra for a family of equilateral triangles with
 $ {  \pi \over 2}  \leq  \omega  \leq  { 2 \pi \over 3}$ . Full points indicate the spectra
of the tiling triangles. Region I corresponds to the neighborhood of tiling triangles,
while II and III denote intermediate regions (see text).} }} }

\pagebreak

\vskip .1cm
\centerline{ { \psfig{figure=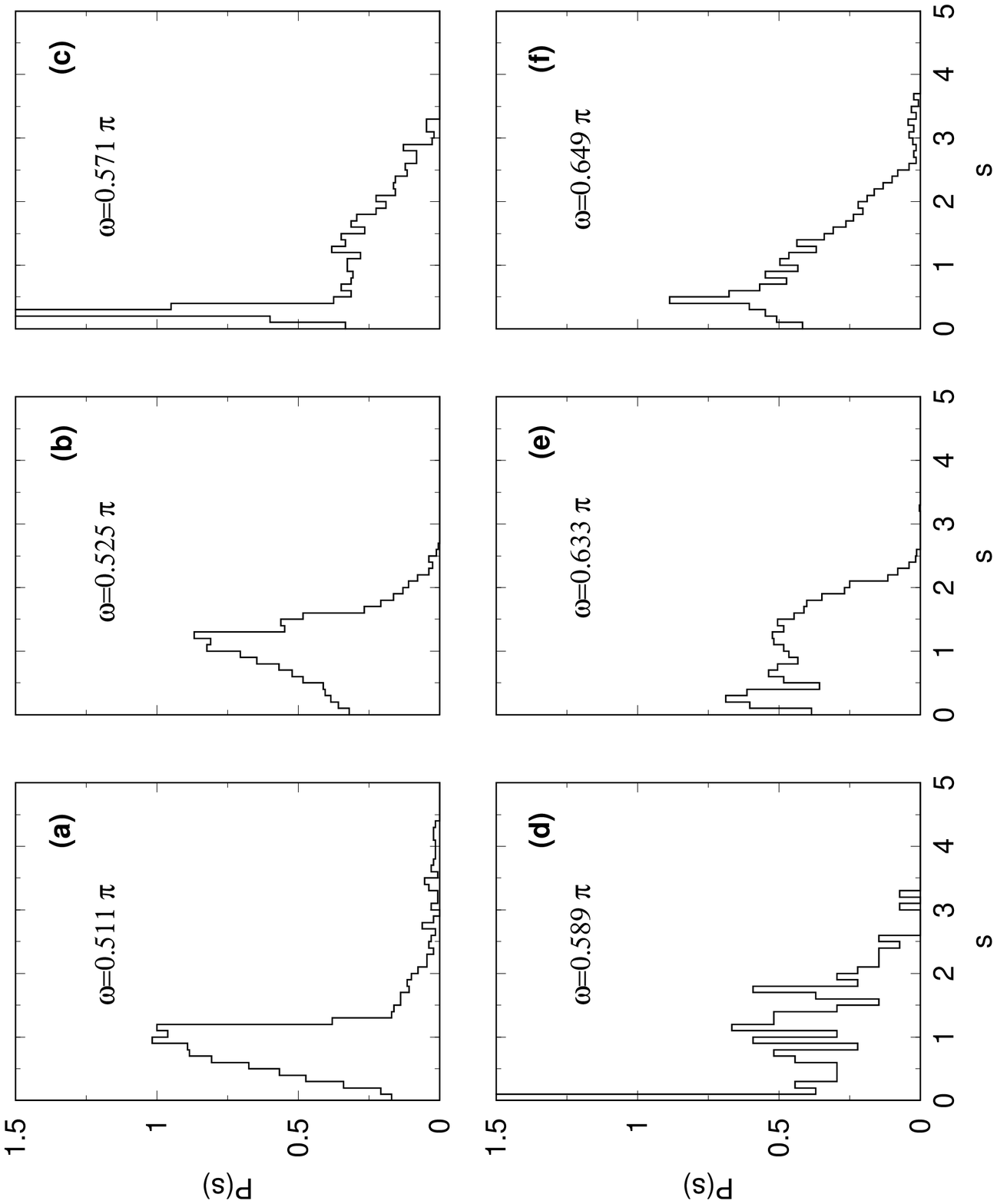,width=20.cm,height=20.cm,angle=0} } }
\vskip .5cm
\centerline{ \parbox{15cm}
{{\normalsize {Fig.3  Level spacing distributions for 6 generic triangles .} }} }

\pagebreak

\vskip .1cm
\centerline{ { \psfig{figure=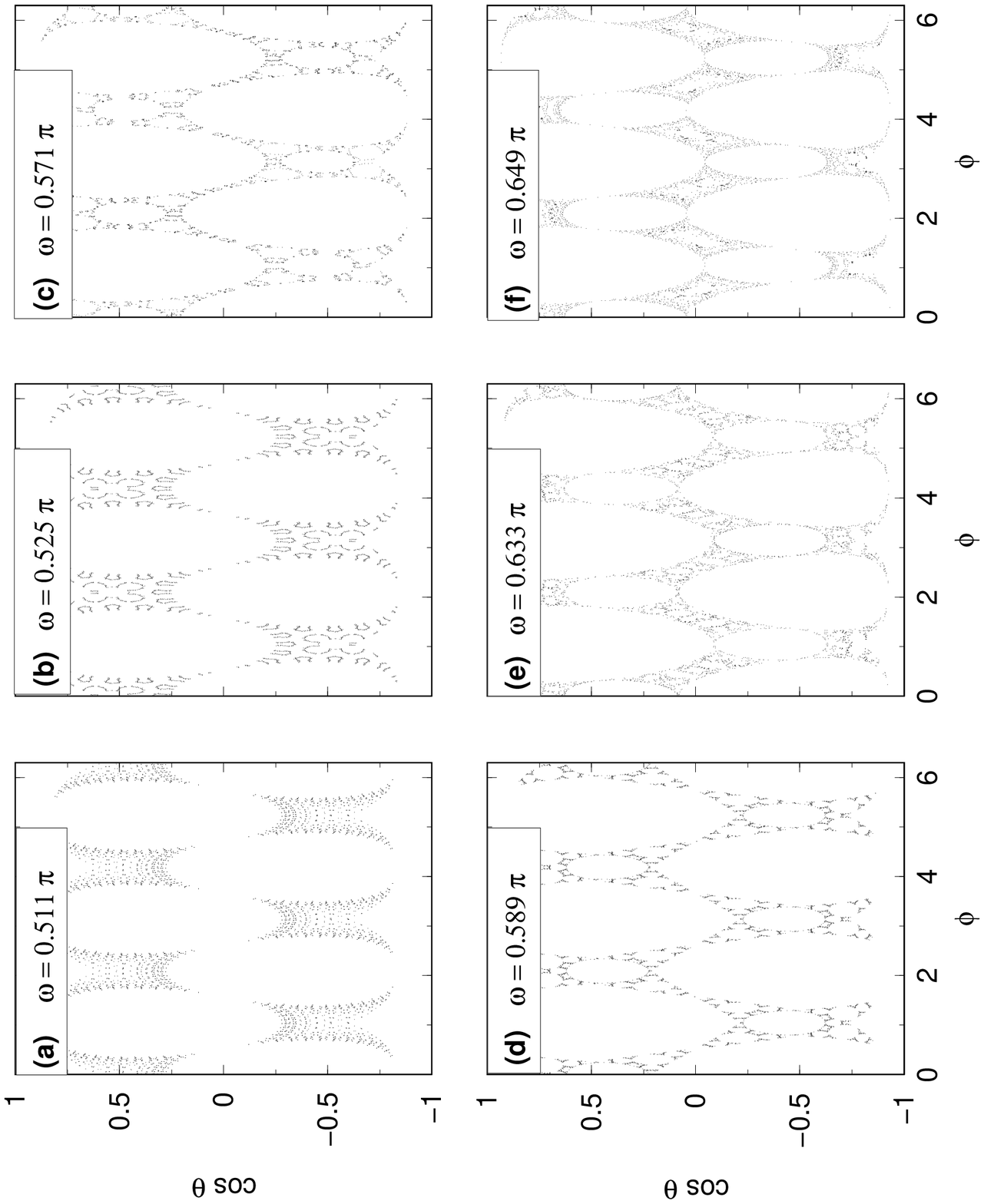,width=20.cm,height=20.cm,angle=0} } }
\vskip .5cm
\centerline{ \parbox{15cm}
{{\normalsize {Fig.4  Phase space plots for 6 generic triangles.} }} }

\pagebreak

\vskip .1cm
\centerline{ { \psfig{figure=pgplot.eps,width=10.cm,height=20.cm,angle=0} } }
\vskip .5cm
\centerline{ \parbox{15cm}
{{\normalsize {Fig.8  Semiclassical representation of three types of eigenfunctions
in the triangle with $ \omega = 0.649 \pi $ .} }} }

\pagebreak

\begin{table}
\caption{\label{ta_1}First levels of the spectrum of tiling triangles with $ \omega=
{ \pi \over 2} $ and $ \omega=
{2 \pi \over 3} $. The energy of each level is $ E = l (l+1) $ and its degeneracy
$ \lambda_l $ .}

\begin{center}

\begin{tabular}{|c|c|c|c|c|}\cline{1-2} \cline{4-5}
\multicolumn{2}{|c|}{$ \omega=
{ \pi \over 2} $}& &\multicolumn{2}{|c|}{$ \omega=
{ 2 \pi \over 3} $} \\ \cline{1-2} \cline{4-5}
\hspace{1cm} l \hspace{1cm}  &  $\hspace{1cm} \lambda_l \hspace{1cm}$ & \hspace{2cm}
&\hspace{1cm} l \hspace{1cm} & \hspace{1cm} $ \lambda_l $ \hspace{1cm} \\
\cline{1-2} \cline{4-5}
 9,13,15,17,19 & 1 &   &  6,9,10,12,13,14,15,16,17 & 1 \\
 21 & 2 &   &  18 & 2 \\
 23 & 1 &   & 19,20 & 1 \\
 25,27,29,31 & 2 &   & 21,22 & 2 \\
 33 & 3 &   & 23 & 1 \\
 35 & 2 &   & 24,25,26,27,28,29 & 2 \\
 37,39,41,43 & 3 &   & 30 & 3 \\
 45 & 4 &   & 31,32 & 2 \\
 47& 3 &   & 33,34 & 3 \\
 49,51,53,55 & 4 &   & 35 & 2 \\

57& 5 &   & 36,37,38,39,40,41 & 3 \\
59& 4 &   & 42 & 4 \\
61,63,65,67,69,71 & 5 &   & 43,44 & 3 \\
73& 6 &   & 45,46 & 4 \\

\cline{1-2} \cline{4-5}
\end{tabular}

\end{center}

\end{table}


\begin{thebibliography}{99}

\bibitem{spina}
Spina M.E and Saraceno M. 1999 {\it J.Phys.A} {\bf 32} 7803


\bibitem{gut}
Gutkin E. 1986 {\it Physica D} {\bf 2} 495

\bibitem{bohi}
Bohigas O.1991  {\it Proceedings of the 1989 Les Houches Summer School on
'Chaos and Quantum Physics'} ed. M.J.Giannoni et al  (North Holland) p 331

\bibitem{berob}
Berry M.V. and Robnik M. 1984 {\it J.Phys.A} {\bf 17} 2413



\bibitem{bogo}
Bogomolny E.B., Gerland U. and Schmit C. 1999 {\it Phys.Rev. E}{\bf 59} R1315


\bibitem{schmi}
Schmit C. 1991  {\it Proceedings of the 1989 Les Houches Summer School on
'Chaos and Quantum Physics'} ed. M.J.Giannoni et al  (North Holland) p 331


\bibitem{table}
Koster J.F.,Dimmock J.O.,Wheeler R.G. and Statz  H. 1963
 {\it Properties of the Thirty-two Point Groups} ( MIT : Cambridge)


\bibitem{itzy}
Itzykson C. and Luck J.M. 1986 {\it J.Phys.A} {\bf 19} 211


\bibitem{berryta}
Berry M.V. and Tabor M. 1977 {\it Proc. R. Soc. Lond. A} {\bf 356} 375


\bibitem{pinsky}
Pinsky M. 1980 {\it SIAM J. Math. Anal.} {\bf 11} 819


\bibitem{berrytri}
Berry M.V. 1981 {\it Ann. Phys. (NY)} {\bf 131} 1633


\bibitem{berrydia}
Berry M.V. and Wilkinson M. 1984 {\it  Proc. R. Soc. Lond. A} {\bf 392} 15


\bibitem{scaling}
Vergini E. and Saraceno M. 1995 {\it  Phys.Rev.E} {\bf 52} 2204



\end{thebibliography}
\end{document}